\documentclass[
reprint,
superscriptaddress,
 amsmath,amssymb,
 aps,
prl,
floatfix,
]{revtex4-1}

\usepackage[english]{babel} 
\usepackage[utf8x]{inputenc} 
\usepackage{microtype} 

\usepackage{graphicx}
\usepackage{bm}
\usepackage{braket}
\usepackage{xr-hyper}
\usepackage{booktabs}
\usepackage[hidelinks]{hyperref}
\usepackage{fixme}

\usepackage[capitalize]{cleveref}
	
\newcommand{\psiin}{\psi_{I}}
\newcommand{\psic}[1]{\psi_{C_{#1}}}
\newcommand{\initial}{\ket{i}}

\newcommand{\HrTwo}{\hat{H}}

\newcommand{\HrN}{\hat{H}}

\newcommand{\jx}[1]{J^x_{#1}}
\newcommand{\jz}[1]{J^z_{#1}}
\newcommand{\js}[1]{J_{S}}
\newcommand{\hc}{\text{h.c.}}

\renewcommand{\i}{I}
\renewcommand{\o}[1]{#1}
\renewcommand{\c}[1]{C_{#1}}
\renewcommand{\b}[1]{B_{#1}}

\begin{document}

\title{A coherent router for quantum networks}

\author{K. S. Christensen}
	\email{kaspersangild@phys.au.dk}
	\affiliation{Department of Physics and Astronomy, Aarhus University, DK-8000 Aarhus C, Denmark.}
\author{S. E. Rasmussen}
	\email{stig@phys.au.dk}
	\affiliation{Department of Physics and Astronomy, Aarhus University, DK-8000 Aarhus C, Denmark.}
\author{D. Petrosyan}
	\email{dap@iesl.forth.gr}
	\affiliation{Institute of Electronic Structure and Laser, FORTH, GR-71110 Heraklion, Greece.}
	\affiliation{A. Alikhanyan National Laboratory, 0036 Yerevan, Armenia.}
\author{N. T. Zinner}%
	\email{zinner@phys.au.dk}
	\affiliation{Department of Physics and Astronomy, Aarhus University, DK-8000 Aarhus C, Denmark.}
	\affiliation{Aarhus Institute of Advanced Studies, Aarhus University, DK-8000 Aarhus C, Denmark.}
\date{\today}

\begin{abstract}
Scalable quantum information processing will require quantum networks of qubits with the ability to coherently transfer quantum states between the desired sender and receiver nodes. Here we propose a scheme to implement a quantum router that can direct quantum states from an input qubit to a preselected output qubit. The path taken by the transferred quantum state is controlled by the state of one or more ancilla qubits. This enables both directed transport between a sender and a number of receiver nodes, and generation of distributed entanglement in the network. We demonstrate the general idea using a two-output setup and discuss how the quantum routing may be expanded to several outputs. We also present a possible realization of our ideas with superconducting circuits.
\end{abstract}

\maketitle

\paragraph{Introduction}\label{sec:introduction}

The transfer of quantum information between different quantum processing units will be an integral part of possible future quantum technology.
While photons will play the decisive role for long-range transfer \cite{cirac1997,ursin2007,kimble2008}, the short-range transport of quantum states is more likely to be accomplished via stationary information channels such as chains and networks of coupled qubits \cite{kielpinski2002,bose2007,kay2010}.
Since the seminal work of Bose~\cite{bose2003}, many studies have explored how to accomplish high-fidelity transfer of quantum states through a spin or qubit network \cite{nikolopoulos2004,subrah2004,christandl2004,albanese2004,wojcik2005,campos2007,difranco2008,franco2008,wu2009,yao2011,kay2011,appollaro2012,Petrosyan2010,Nikolopoulos_book_2014}.

State transfer protocols in such networks typically rely on tuning nearest-neighbor couplings and local fields, either statically or dynamically, in order to maximize the fidelity of moving a quantum state across the network in minimum time.
Controlling the individual qubit energies is usually done with the external classical fields, while couplings between the qubits are tuned via judicious engineering of the inter-qubit interactions \cite{benjamin2003,banchi2011}.  

Since a larger quantum processing unit is likely to consists of several smaller devices or subprocessors, it is crucial to have a quantum routing system for selective high-fidelity state transfer and entanglement sharing between a sender and a distinct receiver in a network. This issue has previously been considered in several  different contexts, including coupled harmonic systems \cite{plenio2004}, external flux threading \cite{bose2005}, local field adjustments in spin systems \cite{wojcik2007,facer2008,nikolopoulos2008,chudzicki2010,pemberton2010,yung2011}, using local periodic field modulation \cite{zueco2009} to manipulate tunneling rates \cite{grossmann1991,grifoni1998,dellavalle2007,creffield2007}, and using optimal control techniques at local sites \cite{pemberton2010}.
The common theme of all of these previous proposals is that they require a considerable amount of careful external control in order to perform the routing of quantum states and entanglement.

In the present work we propose to tune the coupling between the input and the desired output qubits using ancilla qubits. The internal state of the ancilla qubit controls the the direction of the quantum state transfer, serving thus as a quantum router.

The great advantage of our scheme is that the ancilla qubits may be in superposition or entangled states, allowing the router to sent the quantum states into a superposition of different directions. Hence, the process of routing is done in a completely quantum mechanical manner. In combination with, e.g., a set of controllable swapping gates \cite{marchukov2016,rasmussen2019}, quantum routers may be a starting point for constructing physical quantum processing devices analogously to classical circuit designs. 

We first discuss the simplest realization of the router, with just two output qubits. We then describe a router for more than two output qubits. Finally, we propose a concrete realization of a quantum router using superconducting circuits \cite{you2005,devoret2013}. The qubit model used here is general and our routing scheme can also be implemented in numerous other platforms.

\paragraph{Router with two outputs}\label{sec:router_with_two_outputs}
To illustrate the dynamics of the router, we start by considering the router with two output qubits. 
The most elementary quantum router consists of four qubits: The input qubit, the two output qubits and an ancilla qubit that controls the direction of the state transfer from the input to the desired output. We initialize the two output qubits in their ground state $\ket{0}$, while the input and control qubits are initialized in states $\ket{\psiin}$ and $\ket{\psic{}}$ respectively. We write the initial state of the combined system as $\initial=\ket{\psiin{}}\ket{00}\ket{\psic{}}$. The router is then constructed in such a way that if the control qubit is in state $\ket{\psic{}}=\ket{0}$ the input state is moved to the first output qubit, and if the control is in state $\ket{\psic{}}=\ket{1}$ the input state is moved to the second output qubit.
\begin{equation}
	\begin{split}
		\ket{\psiin{}}\ket{00}\ket{0} &\rightarrow \ket{0}\ket{\psiin{}0}\ket{0}, \\
		\ket{\psiin{}}\ket{00}\ket{1} &\rightarrow \ket{0}\ket{0\psiin{}}\ket{1}.
	\end{split}
\end{equation}
In general, if the control qubit is in a superposition state $\ket{\psic{}}=\alpha\ket{0}+\beta \ket{1}$, where $\left|\alpha\right|^2+\left|\beta\right|^2=1$, we have
\begin{equation}\label{eq:routerLinear}
	\ket{\psiin{}}\ket{00}\ket{\psic{}}\rightarrow
	\alpha \ket{0}\ket{\psiin{}0}\ket{0}+ \beta \ket{0}\ket{0\psiin{}}\ket{1}.
\end{equation}
This creates entanglement between the control qubit and the output qubits. Entanglement is a crucial resource in many quantum algorithms and we will show below how the router can be modified such that different types of entanglement are achieved.

\begin{figure}
	\centering
	\includegraphics[width=\columnwidth]{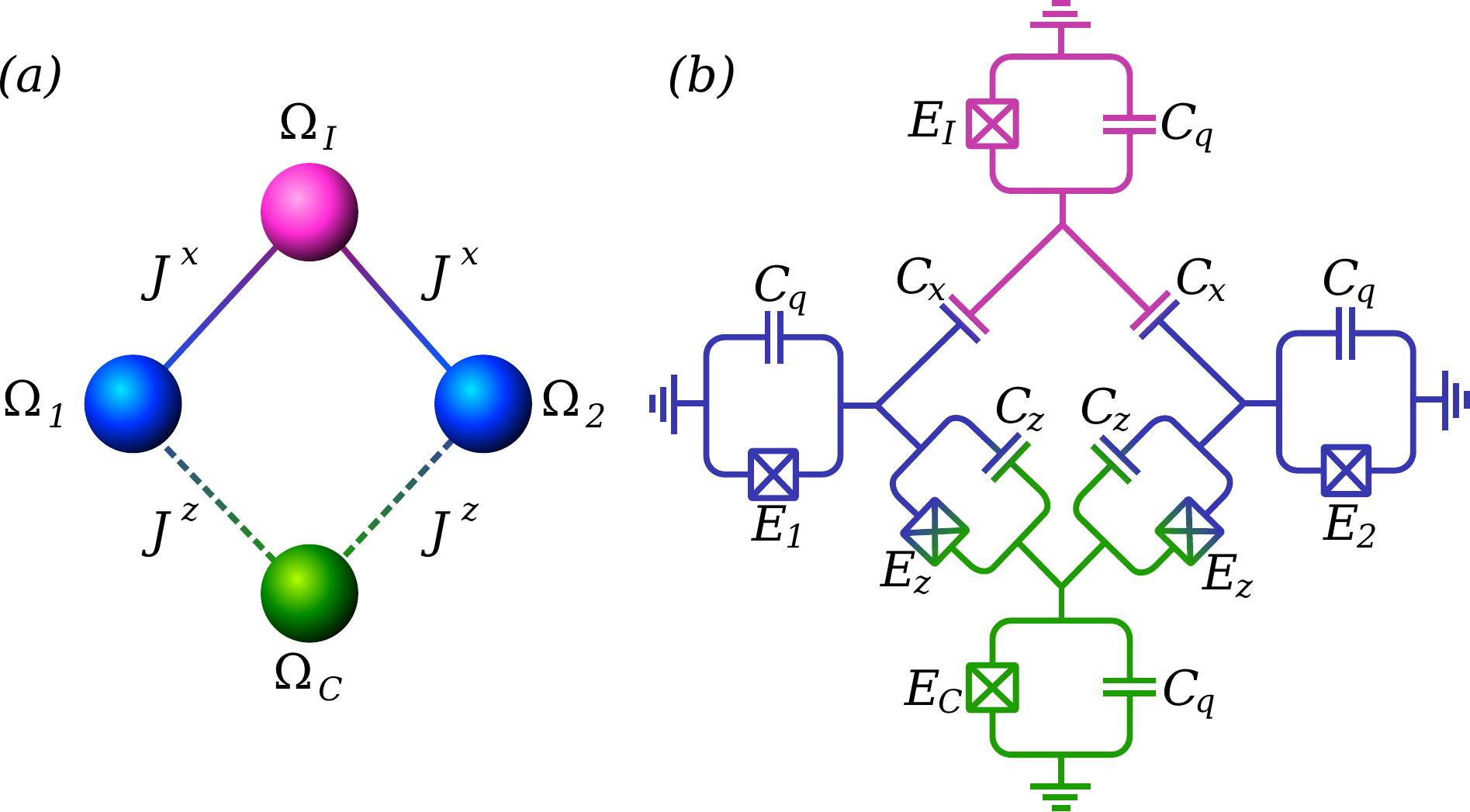}
	\caption{\textbf{(a)} Schematic illustration of the router system with two output qubits. The solid lines represent transverse $XX$-type couplings and the dashed lines represent longitudinal $ZZ$-type couplings. The purple sphere represents the input qubit, the blue spheres represent the output qubits, and the green sphere is the ancilla qubit. Depending on the state of the ancilla qubit, the state of the input qubit is sent to either first or second output qubit, or their superposition. \textbf{(b)} Possible circuit implementation. The superconducting circuit consists of four transmon qubits connected in a square. Two parallel lines indicates capacitors, while the crossed boxes indicate Josephson junctions. The different parts of the system are colored according to their role, as per (a).}
	\label{fig:two_output_router} 
\end{figure}

\Cref{fig:two_output_router}(a) illustrates the system. The Hamiltonian of the quantum router can be written as
\begin{equation}
	\begin{split}
		\HrTwo{} =&  -\frac{\Delta_{\o{1}}}{2}\sigma^z_{\o{1}}-\frac{\Delta_{\o{2}}}{2}\sigma^z_{\o{2}}+\jz{}\left( \sigma^z_{\o{1}} + \sigma^z_{\o{2}}   \right)\sigma_{\c{}}^z \\
		&+ \frac{\jx{}}{2}  \left[ \sigma^x_{\i} \left( \sigma^x_{\o{1}} + \sigma^x_{\o{2}} \right)+\sigma^y_{\i} \left( \sigma^y_{\o{1}} + \sigma^y_{\o{2}} \right) \right],
	\end{split}
  \label{eq:hamil_two_output}
\end{equation}
where $\sigma^x = \ket{0}\bra{1} + \ket{1}\bra{0}$, $\sigma^y = -i\ket{0}\bra{1} + i\ket{1}\bra{0}$ and $\sigma^z = \ket{0}\bra{0} - \ket{1}\bra{1}$ are the Pauli spin operators in the computational basis $\{\ket{0},\ket{1}\}$ of the qubits.  The subscript $\i{}$ indicates the input qubit, while subscripts $\o{1}$ and $\o{2}$ indicate the output qubits, and $\c{}$ the control qubit. The $\ket{0}$-$\ket{1}$ transition frequencies of the output qubits (relative to that of the input qubit) are $\Delta_{\o{1},\o{2}}$, and the transverse and longitudinal coupling strengths are denoted as $\jx{}$ and $\jz{}$ respectively. The first interaction term with strength $ \jz{} $ enables the control qubit to shift the frequencies of the two  output qubits. The second interaction term has strength $\jx{}$ and transversely couples the input qubit to the output qubits. This allows the input qubit to swap an excitation with an output qubit, if their frequencies are resonant. We require the energy shift due to the interaction with the control qubit to be much larger than the transverse coupling $\jz{} \gg \jx{}$.

We assume that the transition frequencies of the output qubits can be independently tuned. Depending on the state of the control qubit, the router should send the state of the input qubit to one of the output qubits. To realize this behavior, we set the detunings as
\begin{equation}\label{eq:detuning_requirement_two_outputs}
 	\Delta_{\o{1}}=-\Delta_{\o{2}}=2\jz{}.
\end{equation}
The diagonal part of the Hamiltonian in \cref{eq:hamil_two_output} then becomes
\begin{equation}\label{eq:hamil_diagonal_part}
	\begin{split}
    \hat{H}_\text{diag} =
    \begin{cases}
      0\sigma^z_{\o{1}}+2\jz{}\sigma^z_{\o{2}} & \text{for } \langle \sigma^z_{\c{}}\rangle = 1\\
      -2\jz{}\sigma^z_{\o{1}}+0\sigma^z_{\o{2}} & \text{for } \langle \sigma^z_{\c{}}\rangle = -1.
    \end{cases}
	\end{split}
\end{equation}
When the control qubit is in the state $\ket{0}$, corresponding to $\langle \sigma^z_{\c{}}\rangle = 1$, the input and the first output qubit are resonant while the second output qubit is detuned.
If the detuning is significantly larger than the transverse coupling strength, i.e. $|4\jz{}/\jx{}|\gg 1$, transfer from the input qubit to second output qubit will be suppressed, while excitations can hop resonantly from the input to the first output qubit. If, on the other hand, the control qubit is in the orthogonal $\ket{1}$ state, the excitation can hop from the input to the second output qubit, while transfer to the first output qubit is suppressed. 

More formally, we may write the Hamiltonian in a frame rotating with its diagonal part as 
\begin{align}\label{eq:hamil_rotating}
		\frac{\hat{H}_{\text{rot}}}{\jx{}} 
		&= 
		\sigma^-_{\i}\sigma^+_\o{1}e^{2i\jz{}\left(\sigma^z_{\c{}}-1 \right)t}+
		\sigma^-_{\i}\sigma^+_\o{2}e^{2i\jz{}\left(\sigma^z_{\c{}}+1 \right)t}
		+\hc\nonumber\\
		&\approx   
		\sigma^-_{\i}\sigma^+_\o{1}\ket{0_{\c{}}}\bra{0_{\c{}}}+
		\sigma^-_{\i}\sigma^+_\o{2}\ket{1_{\c{}}}\bra{1_{\c{}}}
		+\hc,
\end{align}
where we have used the rotating wave approximation in conjunction with the assumption $|4\jz{}/\jx{}|\gg 1$ in order to obtain the final expression. At time $T = \pi/(2\jx{})$ the transfer is complete and the transformation is described by the unitary operator
\begin{equation}\label{eq:ideal_unitary_two_outputs}
\begin{aligned}
    \hat{U}_T = \exp\Big\lbrace
    &-i \frac{\pi}{2}  \big(     \sigma^-_{\i}\sigma^+_\o{1}\ket{0_{\c{}}}\bra{0_{\c{}}}  \\
    &+  \sigma^-_{\i}\sigma^+_\o{2}\ket{1_{\c{}}}\bra{1_{\c{}}}+
    \hc
    \big) \Big\rbrace.
    \end{aligned}
\end{equation}
Note that this unitary transformation is indeed capable of creating entanglement when the control qubit is in a superposition state, as in \cref{eq:routerLinear}.

\begin{figure}
	\centering
	\includegraphics[width = 0.9\columnwidth]{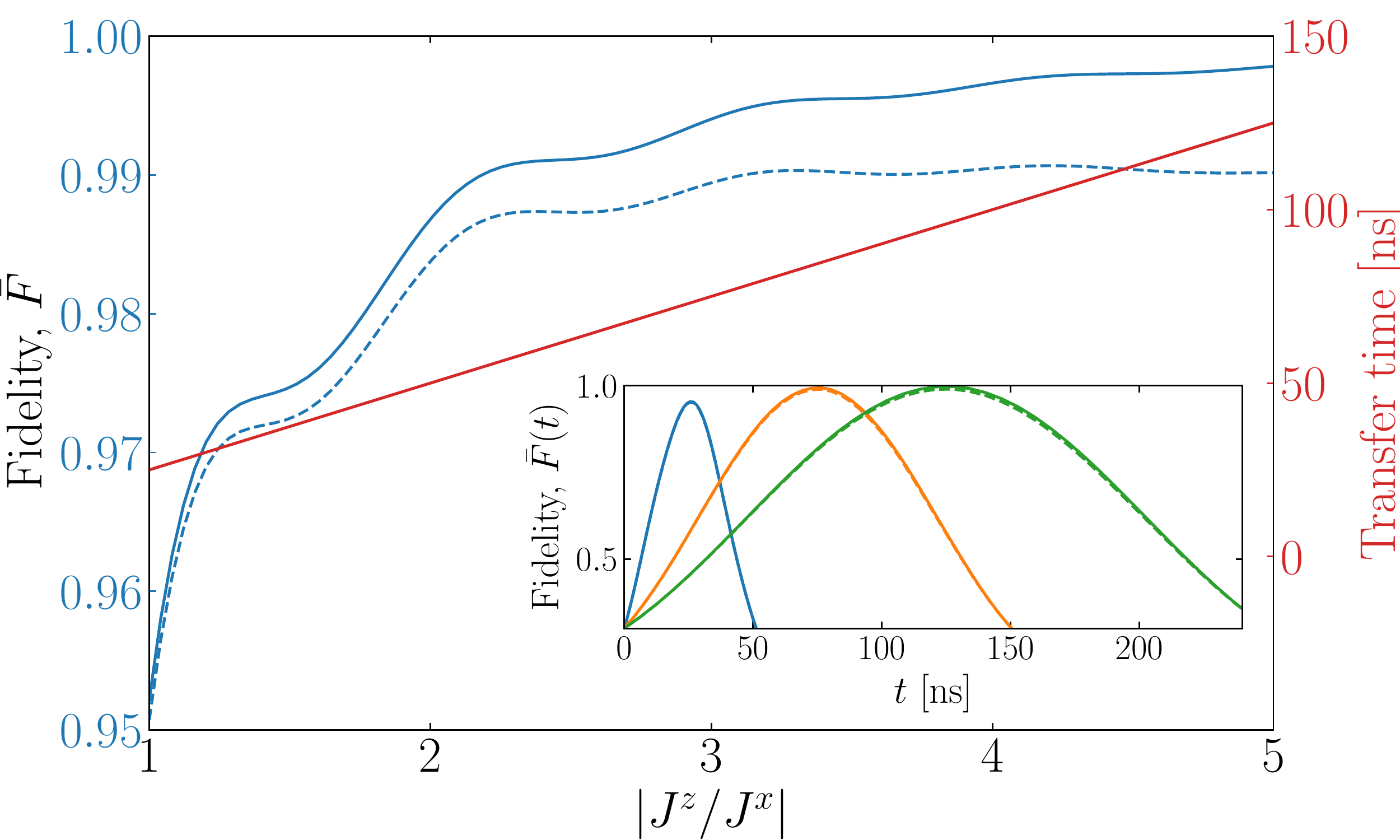}
	\caption{Average fidelity of the two-output router as a function of the coupling ratio. The blue solid line shows the average fidelity at time $ t=\pi/(2J^x) $. The dashed lines are the average fidelity with relaxation and coherence time of $T_1=T_2 = 30\mu$s. The red line is the corresponding transfer time. In all calculations we have used $J^z/(2\pi) = 10$MHz. Insert:
	Time dependence of the state transfer of the two-output router for $\jz{}/\jx{} = 1$ (blue), $\jz{}/\jx{} = 3$ (yellow) and $\jz{}/\jx{} = 5$ (green).}
	\label{fig:fid_vs_jzjx}
\end{figure}

To characterize the performance of the quantum router, we calculate the average process fidelity, defined as \cite{nielsen_chuang_2010,nielsen2002,horodecki1999,schumacher1996}
\begin{equation}\label{eq:av_fidelity_formula}
\bar{F} = \int d\psi\bra{\psi}\hat{U}_T^\dagger\mathcal{E}(\psi)\hat{U}_T\ket{\psi},
\end{equation}
where the integration is performed over the subspace of all possible initial states and $\mathcal{E}$ is the quantum map realized by our system. We initialize the two output qubits in state $\ket{0}$ so the subspace of initial states is spanned by $\left\{	\ket{0}\ket{00}\ket{0},\allowbreak\ket{1}\ket{00}\ket{0},\allowbreak\ket{0}\ket{00}\ket{1},\allowbreak\ket{1}\ket{00}\ket{1} \right\}$. The average fidelity is then calculated with the QuTiP Python toolbox \cite{qutip} using the procedure described in \cite{PEDERSEN2007}. In all calculations, we have $J^z/(2\pi)= 10\text{MHz}$ and the relaxation and decoherence times are $T_1=T_2 = 30\mu$s \cite{Wendin_2017}. In \cref{fig:fid_vs_jzjx} we show the average process fidelity at the transfer time $T=\pi/(2J^x)$. When $|J^x| = |J^z|$ the most detrimental source of error is transfer to the wrong output qubit, since the detuning induced by the control qubit is not large enough to completely suppress the hopping interaction connecting the input and closed output qubits. In this regime, the error due to decoherence is comparatively small, which is due to the fact that the transfer times are shorter for larger $J^x$. For larger values of $|J^z/J^x|$, transfer to the closed output qubit is stronger suppressed, and the average process fidelity approaches unity, if we neglect decoherence. But since the transfer time also increases, decoherence becomes the dominant source of error. With our choice of parameters, the maximum fidelity is $\bar{F}_{max} = 0.9907$ at $|J^z/J^x| = 4.192$. 

\paragraph{Concatenated routers}
\label{sub:concatenated_routers}

\begin{figure}
	\centering
	\includegraphics[width=0.85\columnwidth]{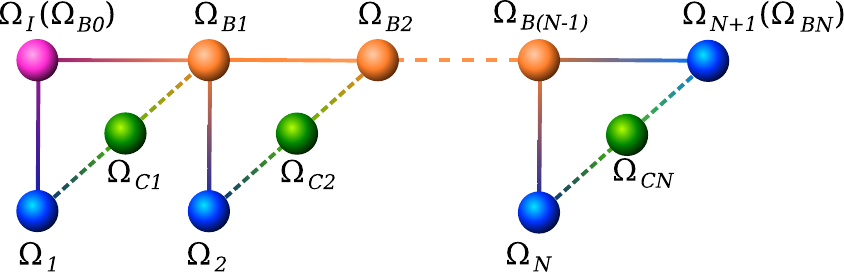}
	\caption{\textbf{(a)} Schematic illustration of a concatenated router with N+1 output ports. The purple sphere is the input qubit, the blue spheres are the output qubits, the green spheres are the control qubits, and the orange spheres are bus qubits, i.e., qubits which act both as input and output qubits.}
	\label{fig:concatenated_router} 
\end{figure}

The number of output ports of the router can be scaled in several ways (see Supplementary material \cite{supplMat}). Here we describe a scheme in which $N$ routers are concatenated as shown on \cref{fig:concatenated_router} where one output of each router serves as the input qubit for the next one. We refer to these qubits as the \emph{bus} qubits. 

The concatenated router operates in (time) steps. \emph{Step 0:} Initialize the input qubit in a given state $\ket{\psi}$. \emph{Step 1:} The state will either move down to the first output or right to the first bus qubit, depending on the state of the first control qubit. After the state have been transferred, i.e., at $t=T$, the input qubit is closed by detuning it from the bus qubit. \emph{Step 2:} The state moves either down to the second output qubit or continues right to the second bus qubit, depending on the state of the second control qubit. \emph{Step 3:} Detune the second bus qubit from the first bus qubit at time $t=2T$. The procedure proceeds as above until the state moves down into one of the $N$ output qubits or it arrives at the last output qubit $N+1$.

This process can be expressed through a Hamiltonian with time-dependent detunings. The static part of the Hamiltonian is
\begin{align}
	\HrN{} =& J^z\sum_{i=1}^{N}\left[\sigma_{\o{i}}^z(\sigma_{\c{i}}^z + 1) + \sigma_{\b{i}}(\sigma_{\c{i}}^z - 1)\right] \\
	&+\frac{J^x}{2} \sum_{i=1}^{N}\left[ \sigma_{\b{i-1}}^x(\sigma_{\o{i}}^x+\sigma_{\b{i}}^x) + \sigma_{\b{i-1}}^y(\sigma_{\o{i}}^y+\sigma_{\b{i}}^y)\right], \nonumber
\end{align}
where we denote the bus' qubits $\sigma_{\b{}}^{x,y,z}$, and the zero'th bus qubit is the input qubit, while the $N$'th bus qubit is the final output qubit. The time-dependent part of the Hamiltonian is
\begin{equation}
\begin{aligned}
	\HrN{}(t) =& \Delta\sum_{i=1}^{N} \left[ \sigma_{\b{i-1}}^z + \sigma_{\b{i}}^z + \sum_{j=1}^{i}\sigma_{\o{j}}^z \right]\\
	&\qquad\times \left[\theta(t - (i-1)T) - \theta(t+iT)\right],
\end{aligned}
\end{equation}
where $\Delta$ is the detuning, and $\theta(t)$ is the Heaviside step function. For the process to function properly, we must ensure that $|\Delta| \gg |J^x|$ and $|2J^z\pm \Delta| \gg |J^x|$. Thus, an excitation starting in the input qubit will move down the chain of bus qubits in discrete time-steps until it encounters a control qubit in state $ \ket{0} $ and moves to the associated output qubit, where it will remain for the rest of the process. 

\paragraph{Implementation using superconducting circuits}
\label{sub:circuit_implementation}
\begin{table}
\caption{Physical parameters used for our example implementation of the quantum router.}
\label{tab:circ_params}
\begin{tabular}{rrrrrrrr}
\toprule
$ \frac{C_{q}}{\text{fF}}$ & $\frac{C_{z}}{ \text{fF}}$ &  $ \frac{C_{x}}{\text{fF}}$ &  $\frac{E_{I}}{2 \pi\text{GHz}}$ &  $\frac{E_{1}}{2 \pi\text{GHz}}$ &  $\frac{E_{2}}{2\pi\text{GHz}}$ &  $\frac{E_{C}}{2\pi\text{GHz}}$ &  $\frac{E_{z}}{2\pi\text{GHz}}$ \\
\midrule
 80.0 &                         13.7 &                     0.082 &                         19.52 &                         19.22 &                         19.52 &                         38.74 &                           3.46 \\
\bottomrule
\end{tabular}
\end{table}
Superconducting circuits present a promising platform to implement the quantum router. Specifically, we propose an implementation using transmon qubit architecture as shown in \cref{fig:two_output_router}(b). The circuit consists of four transmon qubits, each of which can be made flux-tunable by substituting the Josephson Junction with a SQUID. The two output qubits (blue) are coupled to the control qubit (green) each through a Josephson junction and a capacitor in parallel. The nonlinearity of this Josephson Junction provides the main mechanism behind the $ZZ$-type coupling between outputs and control.  A tunable version of this coupler has been investigated experimentally, and it has been shown that the transversal coupling could be made negligible compared to the longitudinal coupling \cite{Kounalakis2018}. In our scheme, the transverse coupling between the output qubits and control is much smaller than their relative detuning such that there will be no exchange of excitations between them.

By using second order perturbation theory, we derive an effective Hamiltonian for the circuit (see supplementary material \cite{supplMat})
\begin{equation}
	\begin{split}
	\hat{H}_{\textbf{eff}}^{\textbf{circ}} = &  -\frac{\Delta_{\o{1}}}{2}\sigma^z_{\o{1}}-\frac{\Delta_{\o{2}}}{2}\sigma^z_{\o{2}}+\jz{}\left( \sigma^z_{\o{1}} + \sigma^z_{\o{2}}   \right)\sigma_{\c{}}^z \\
	&+ \frac{\jx{}}{2}  \left[ \sigma^x_{\i} \left( \sigma^x_{\o{1}} + \sigma^x_{\o{2}} \right)+\sigma^y_{\i} \left( \sigma^y_{\o{1}} + \sigma^y_{\o{2}} \right) \right]\\
	&+\left(\frac{J^x_{12}}{2} + \frac{J^{xz}_{12}}{2}\sigma^z_{\c{}}\right)\left(  \sigma^x_{\o{1}} \sigma^x_{\o{2}} +\sigma^y_{\o{1}}  \sigma^y_{\o{2}}\right),
	\end{split}
\end{equation} 
where the three last terms arise from the second order interactions between outputs and control. A numerical modeling of the full circuit Hamiltonian with the parameters shown in \cref{tab:circ_params} gives us a longitudinal coupling of $J^z = -9.95\cdot(2\pi\text{MHz})$ and a transversal input/output coupling of $J^x=2.78(2\pi \text{MHz})$ with appropriate detunings of the outputs, as explained in the introduction. The three remaining couplings are all much smaller than $J^x$ and can thus be neglected for our purposes. 

\paragraph{Conclusion}
\label{sec:conclusion}
We have presented a simple implementation of a quantum router with quantum control, and analyzed it analytically and numerically. By utilizing a relatively strong interaction with a control qubit, state transfer to the undesired output qubit can be suppressed, and we achieve selective transfer fidelity above 0.99, even when including the effects of dephasing and relaxation. We have also presented a scalable scheme that can extend the router to an arbitrary number of outputs. Finally, we have presented a possible realization of the router using a superconducting circuit. Our simple routing scheme is capable of distributing entanglement between distant qubits which is highly useful for short-range (on-chip) quantum communications. 

\begin{acknowledgments}
	The authors would like to thank T. Bækkegaard, L. B. Kristensen, and N. J. S. Loft for discussion on different aspects of the work.
	This work is supported by the Danish Council for Independent Research and the Carlsberg Foundation.
	D.P. is partially supported by the HELLAS-CH (MIS Grant No. 5002735), and is grateful to the Aarhus Institute of Advanced Studies for hospitality.
\end{acknowledgments}


%

\end{document}


\title{Supplementary Material for "A coherent router for quantum networks"}

\author{Kasper Sangild Christensen}
\email{kaspersangild@phys.au.dk}
\affiliation{Department of Physics and Astronomy, Aarhus University, DK-8000 Aarhus C, Denmark.}
\author{Stig Elkjær Rasmussen}
\email{stig@phys.au.dk}
\affiliation{Department of Physics and Astronomy, Aarhus University, DK-8000 Aarhus C, Denmark.}
\author{David Petrosyan}
\email{dap@iesl.forth.gr}
\affiliation{Institute of Electronic Structure and Laser, FORTH, GR-71110 Heraklion, Greece.}
\author{Nikolaj Thomas Zinner}%
\email{zinner@phys.au.dk}
\affiliation{Department of Physics and Astronomy, Aarhus University, DK-8000 Aarhus C, Denmark.}
\affiliation{Aarhus Institute of Advanced Studies, Aarhus University, DK-8000 Aarhus C, Denmark.}
\date{\today}
	\maketitle
	
\section{Superconducting circuit implementation of the router}
\label{secSM:implementation_using_superconducting_circuits}

In this section we will show how the Hamiltonian of Eq. (3) in  the main text can be realized using the circuit in Fig. 1(b). We use the procedure presented by Refs. \cite{Devoret1997,Devoret2017}. Since the couplings between the input and output are much smaller than the couplings between the outputs and control, we will analyze an isolated system consisting of the outputs and control before adding in the input qubit. The circuit Hamiltonian can then be written as
\begin{equation}\label{eq:circ_hamil}
H = \frac{1}{2}\vec{q}^T\bm{C}^{-1}\vec{q}
	- E_{J1} \cos{\left (\varphi_{1} \right )}- E_{J2} \cos{\left (\varphi_{2} \right )}- E_{JC} \cos{\left (\varphi_{C} \right )} - E_{Jz1} \cos{\left (\varphi_{1} - \varphi_{C} \right )}  - E_{Jz2} \cos{\left (\varphi_{2} - \varphi_{C} \right )},
\end{equation} 
Here $\{\varphi_i\}_{i\in\{1,2,C\}}$ are phase differences across the Josephson junctions of the respective qubits, and $q_i$ are the conjugate charge operators fulfilling $[\varphi_i,q_j] = i\delta_{ij}$. The capacitance matrix is given explicitly by:
\begin{equation}\label{eq:capacitance_matrix}
	\bm{C}=\left[\begin{matrix}C_{1} + C_{z1} & 0 & - C_{z1}\\0 & C_{2} + C_{z2} & - C_{z2}\\- C_{z1} & - C_{z2} & C_{C} + C_{z1} + C_{z2}\end{matrix}\right].
\end{equation}
For a typical transmon, the charging energy is much smaller than the junction energy and the phase is well localized near the bottom of the potential. This is equivalent to a heavy particle moving near the equilibrium position. We can thus use the fourth order Taylor expansion of the full potential, which allows us to rewrite the Hamiltonian as
\begin{equation}\label{eq:hamiltonian_taylor_expansion}
    \begin{split}
        \hat{H} =&\sum_{i\in\{1,2,C\}}\left[\sqrt{8\widetilde{E}_{Ji}E_{Ci}}\hat{b}_{i}^{\dagger}\hat{b}_{i}-\frac{E_{Ci}}{12}\left(\hat{b}_i^\dagger+\hat{b}_i\right)^4\right] - \sum_{i\neq j \in \{1,2,C\}}\frac{\left(\bm{C}^{-1}\right)_{i,j}}{4\sqrt{\zeta_i\zeta_j}}\left(\hat{b}_i^\dagger-\hat{b}_i\right)\left(\hat{b}_j^\dagger-\hat{b}_j\right)\\
        &+\sum_{i\in\{1,2\}} E_{Jzi}
        \left[
        \frac{\zeta_C\sqrt{\zeta_i\zeta_C}\left(\hat{b}_i^\dagger
            +\hat{b}_i\right)\left(\hat{b}_C^\dagger+\hat{b}_C\right)^3+
            \zeta_i\sqrt{\zeta_i\zeta_C}\left(\hat{b}_i^\dagger+\hat{b}_i\right)^3\left(\hat{b}_C^\dagger+\hat{b}_C\right)}{24}
        -\frac{\sqrt{\zeta_i\zeta_C}\left(\hat{b}_i^\dagger+\hat{b}_i\right)\left(\hat{b}_C^\dagger+\hat{b}_C\right)}{2}\right]\\
        &-\sum_{i\in\{1,2\}} E_{Jzi}\frac{\zeta_i\zeta_C\left(\hat{b}_i^\dagger+\hat{b}_i\right)^2\left(\hat{b}_C^\dagger+\hat{b}_C\right)^2}{16},
    \end{split}
\end{equation}
where we have defined the effective single mode Josephson energies and charging energies
\begin{equation}\label{eq:single_mode_josephson_energies}
	\begin{split}
		&\widetilde{E}_{J1} = E_{J1}+E_{Jz1},\quad
		\widetilde{E}_{J2} = E_{J2}+E_{Jz2},\quad
		\widetilde{E}_{JC} = E_{JC}+E_{Jz1}+E_{Jz2},\\
        &E_{C1} = \frac{\left(\bm{C}^{-1}\right)_{1,1}}{8},\quad
        E_{C2} = \frac{\left(\bm{C}^{-1}\right)_{2,2}}{8},\quad
        E_{CC} = \frac{\left(\bm{C}^{-1}\right)_{3,3}}{8},
	\end{split}
\end{equation}
and the ladder operators
\begin{equation}\label{eq:phi_and_q_using_ladder_ops}
\begin{split}
\hat{\varphi}_i = \sqrt{\frac{\zeta_i}{2}}\left(\hat{b}_i^\dagger+\hat{b}_i\right),\quad \hat{q} = \frac{i}{\sqrt{2\zeta_i}}\left(\hat{b}_i^\dagger-\hat{b}_i\right),
\end{split}
\end{equation}
with impedances $\zeta_i = \sqrt{(\bm{C}^{-1})_{i,i}/\widetilde{E}_{Ji} }$.
Note that even though there is no capacitor between qubits 1 and 2, there is still a capacitative coupling between the two qubits, since $(\bm{C}^{-1})_{1,2}$ is non-zero. The circuit operates in the weak coupling limit $ E_{Jzi} \ll E_{Jj} \forall i,j $ and $ C_{zi} \ll C_{j} \forall i,j $. This allows us to view the system as three harmonic oscillators perturbed by the quadratic and quartic interactions. In addition, we will assume that modes 1 and 2 are very close to resonance such that we can treat their detuning as part of the perturbation. For simplicity, we neglect terms that do not preserve the number of excitations, such as $\hat{b}_1^\dagger\hat{b}_2^\dagger$. Such terms are suppressed by a large energy gap, and thus only give rise to minor corrections. The total Hamiltonian is then the sum of the uncoupled harmonic oscillator Hamiltonian $\hat{H}_0$ and a perturbation $\hat{V}$
\begin{equation}\label{eq:simplified_hamiltonian}
	\begin{split}
        \hat{H}_0 = & \overline{\omega}\left(\hat{n}_1+\hat{n}_2+\hat{n}_C\right),\\
        \hat{V} = 
        &\delta \left(\hat{n}_{2}-\hat{n}_{1}\right)+\Delta \hat{n}_{C}+ 
        \frac{\alpha_{1} \hat{n}_{1} \left(\hat{n}_1-1\right)}{2}+ \frac{\alpha_{2} \hat{n}_{2} \left(\hat{n}_2-1\right)}{2}+ \frac{\alpha_{C} \hat{n}_{C} \left(\hat{n}_C-1\right)}{2}\\
        &+g_{z1} \hat{n}_{1} \hat{n}_{C}+g_{z2} \hat{n}_{2} \hat{n}_{C}+g_{x1} \hat{X}_{1C}+g_{x2} \hat{X}_{2C}+g_{x12} \hat{X}_{12}\\
        &+g_{xz1} \left(\zeta_{1} \left(\hat{X}_{1C} \hat{n}_{1} + \hat{n}_{1} \hat{X}_{1C}\right) + \zeta_{C} \left(\hat{X}_{1C} \hat{n}_{C} + \hat{n}_{C} \hat{X}_{1C}\right)\right)\\
        &+g_{xz2} \left(\zeta_{2} \left(\hat{X}_{2C} \hat{n}_{2} + \hat{n}_{2} \hat{X}_{2C}\right) + \zeta_{C} \left(\hat{X}_{2C} \hat{n}_{C} + \hat{n}_{C} \hat{X}_{2C}\right)\right)
    \end{split}
\end{equation}
The unperturbed qubit frequencies and anharmonicities are then given by
\begin{equation}\label{eq:frequencies}
    \begin{split}
        \omega_{i} &= \sqrt{8\widetilde{E}_{Ji}E_{Ci}}-\alpha_i-\frac{E_{Jzi}\zeta_i\zeta_{C}}{8} \quad i=1,2,\\
        \omega_{C} &= \sqrt{8\widetilde{E}_{JC}E_{CC}}-\alpha_C-\frac{E_{JzC}\left(\zeta_1+\zeta_{2}\right)\zeta_{C}}{8}, \\
        \alpha_i &= -E_{Ci} \quad i = 1,2,C,\\
        \overline{\omega} &= \frac{\omega_1+\omega_2}{2},\\
        \delta &= \frac{\omega_1-\omega_2}{2},\\
        \Delta &= \omega_C-\overline{\omega}.
    \end{split}
\end{equation}
and the coupling strengths are given by
\begin{equation}\label{eq:coupling_strengths}
    \begin{split}
    	g_{zi} &= -\frac{E_{Jzi}\zeta_i\zeta_{C}}{4} \quad i = 1,2,\\
        g_{x12} &= \frac{\left(\bm{C}^{-1}\right)_{1,2}}{2\sqrt{\zeta_{1}\zeta_{2}}},\\
        g_{xi} & = \frac{\left(\bm{C}^{-1}\right)_{i,3}}{2\sqrt{\zeta_{i}\zeta_{C}}}-\frac{E_{Jzi}\sqrt{\zeta_{i}\zeta_{C}}}{2}
        +\frac{E_{Jzi}\left(\zeta_{i}+\zeta_{C}\right)\sqrt{\zeta_{i}\zeta_{C}}}{16}\quad i = 1,2,\\
        g_{xzi} &= \frac{E_{Jzi}\sqrt{\zeta_{i}\zeta_{C}}}{16}\quad i = 1,2.
    \end{split}
\end{equation}
The unperturbed Hamiltonian has a degenerate spectrum with the lowest lying energies being $E_D \in \{0,\Delta,\overline{\omega},\overline{\omega}+\Delta,2\overline{\omega},2\overline{\omega}+\Delta, 3\overline{\omega}+\Delta\}$. If the detuning $\Delta$ is much larger than the transverse coupling between the control and the output qubits, we can ignore the first order excitation swaps between the control and the output. In this case each degenerate subspace is well described by an effective interaction
\begin{equation}\label{eq:eff_ham_general_formula}
	\hat{P}\hat{V}_{eff}\hat{P} = \hat{P}\hat{V}\hat{P}+\hat{P}\hat{V}\hat{Q}\frac{1}{E_D-\hat{Q}\hat{H}_0\hat{Q}}\hat{Q}\hat{V}\hat{P},
\end{equation}
where $\hat{P}$ projects onto the degenerate subspace and $\hat{Q}=1-\hat{P}$ projects onto the orthogonal complement. If the anharmonicity is much larger than the total transverse coupling between qubits 1 and 2, we can justify projecting the final effective Hamiltonian onto the two lowest states of each qubit. In doing so, we find that the effective interaction between the three qubits is given by
\begin{align}\label{eq:eff_spin_model_hamil_Control_and_outputs}
		\begin{split}
		\hat{V}_{eff}= &  -\frac{\Delta_{\o{1}}}{2}\sigma^z_{\o{1}}-\frac{\Delta_{\o{2}}}{2}\sigma^z_{\o{2}}+\jz{}\left( \sigma^z_{\o{1}} + \sigma^z_{\o{2}}   \right)\sigma_{\c{}}^z +\left(\frac{J^x_{12}}{2} + \frac{J^{xz}_{12}}{2}\sigma^z_{\c{}}\right)\left(  \sigma^x_{\o{1}} \sigma^x_{\o{2}} +\sigma^y_{\o{1}}  \sigma^y_{\o{2}}\right).\\
		\end{split}
\end{align}

The qubit frequencies can be calculated from the second order matrix elements found from \cref{eq:eff_ham_general_formula}
\begin{equation}\label{eq:frequencies_explicit}
\begin{split}
		\Delta_1 	=& 	-\delta +           
                        \frac{g_{z1C}}{2}-\frac{\gamma_{1C}(1,1)+\gamma_{1C}(1,3)-\gamma_{1C}(3,1)}{\Delta},\\
		\Delta_2 	=& 		\delta + \frac{g_{z2C}}{2}-\frac{\gamma_{2C}(1,1)+\gamma_{2C}(1,3)-\gamma_{2C}(3,1)}{\Delta},\\
		\Delta_C 	=&		\frac{g_{z1C}+g_{z2C}}{2}+\frac{\gamma_{1C}(1,1)+\gamma_{1C}(1,3)-\gamma_{1C}(3,1)+\gamma_{2C}(1,1)+\gamma_{2C}(1,3)-\gamma_{2C}(3,1)}{\Delta},
\end{split}
\end{equation}
where we have defined $\gamma_{iC}(n,m) = g_{xi}+g_{xzi}\left(n\zeta_{i}+m\zeta_{C}\right)$.

The longitudinal couplings between the control and outputs 1 are
\begin{equation}\label{eq:longitudinal_coupling}
	\begin{split}
		J^z_i =	\frac{g_{z1}}{4}+\frac{\gamma_{1C}(3,1)-\gamma_{1C}(1,3)}{2\Delta}\quad i = 1,2.
	\end{split}
\end{equation}
As described in the main text, the purpose of this longitudinal coupling is to suppress state transfer to the closed output qubit. We thus require this coupling to be significantly larger than the coupling between the input and output qubits.
\subsection{Residual coupling between the outputs}
In \cref{eq:eff_spin_model_hamil_Control_and_outputs} there is an undesired coupling between the two output qubits. The strength of this coupling is given by:
\begin{equation}\label{eq:residual_out-out_coupling}
\begin{split}
		J^x_{12} =& g_{x12}-\frac{\gamma_{1C}(1,3)\gamma_{2C}(1,3)}{\Delta},
				\\
		J^{xz}_{12} =& \frac{\gamma_{1C}(1,3)\gamma_{2C}(1,3)-\gamma_{1C}(1,1)\gamma_{2C}(1,1)}{\Delta}.
\end{split}
\end{equation}
Here we notice an interesting feature that the coupling strength depends on the state of the control qubit. This may be useful in other applications, such as the implementation of controlled three qubit gates, but for our purposes we require this coupling to be as small as possible. If it is much smaller than the longitudinal coupling between the control and outputs, transfer between the two outputs we be suppressed by a detuning much larger than $|J^x_{12}\pm J^{xz}_{12}|$ and can thus be neglected.

\subsection{Coupling to the input qubit}
In our model we couple the input qubit capacitively to the outputs, but one could also couple the qubits through a resonator or transmission line. The only requirement is that it must produce a transverse coupling between the input and outputs. For a small coupling capacitance $C_x$, the coupling strength is given by
\begin{equation}\label{eq:in_out_coupling}
	J^x_{I,j} \approx \frac{C_x}{2\widetilde{C}_I\widetilde{C}_i\sqrt{\zeta_1\zeta_2}}.
\end{equation}

\section{Implementation of three output router}
\label{subSM:implementation_of_three_output_router}
\begin{figure}
	\centering
	\includegraphics[width=0.7\columnwidth]{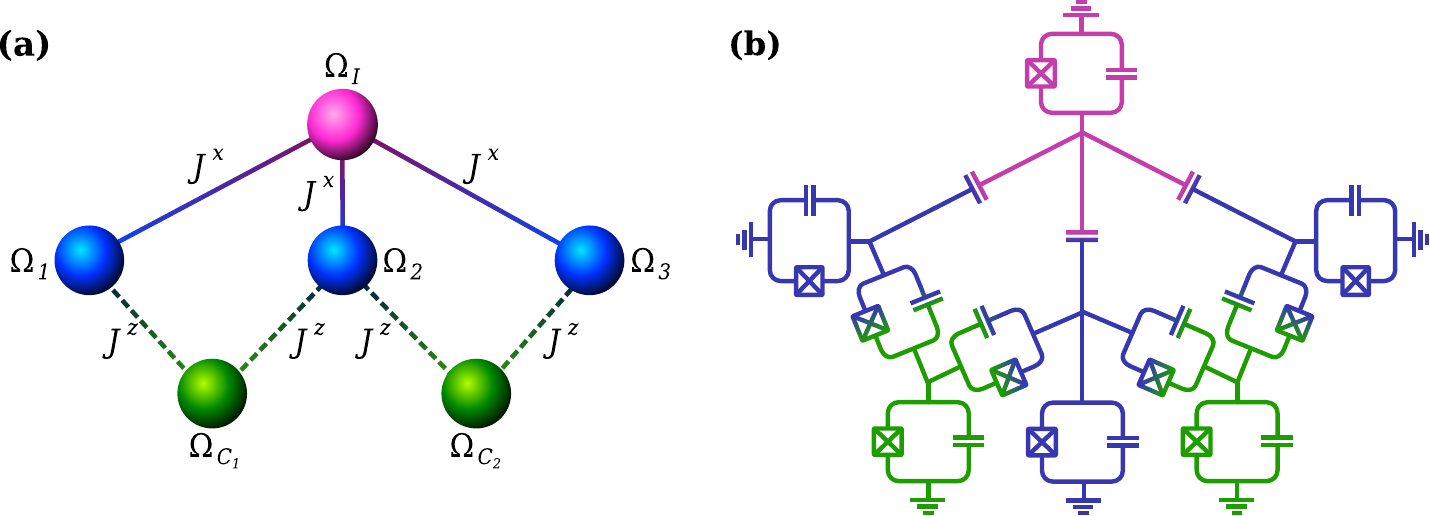}
	\caption{\textbf{(a)} Schematic illustration of the router system with three output qubits. The solid lines represent transverse $XX$-type couplings and the dashed lines represent longitudinal $ZZ$-type couplings. The purple sphere represents the input qubit, the blue spheres represents the output qubits, and the green spheres the control qubits. Depending on the state of the control qubits, the state of the input qubit is sent to one of the output qubits, or a superposition of these. Compared to \cref{fig:two_output_router}(a) an extra output and control qubit have been added. \textbf{(b)} Possible circuit implementation The circuit consists of six connected Transmon qubits according to the scheme illustrated in \textbf{(a)}. The different parts of the two systems are colored according to which part of the system they corresponds to.}
	\label{fig:three_output_router} 
\end{figure}
In the main text, we describe the implementation of a multiple output port router by concatenating the two-port routers. Here we discuss a different implementation of the three output port router. In this section we discus the three output router in detail. A schematic illustration of a three output router is shown in \cref{fig:three_output_router}. The system can be described using a spin model Hamiltonian
\begin{equation}\label{eq:3_out_spin_model}
\begin{split}
	H = &
	-\sum_{i=1}^3\Delta_i\sigma^z_i 
	+ J^z\left(
	\left(\sigma^z_1+\sigma^z_2\right)\sigma^z_{C1}+\left(\sigma^z_2+\sigma^z_3\right)\sigma^z_{C2}
	\right)\\
	& +
	\frac{J^x}{2}\left(\sigma^x_I\left(\sigma^x_1+\sigma^x_2+\sigma^x_3\right) 
	+\sigma^y_I\left(\sigma^y_1+\sigma^y_2+\sigma^y_3\right)\right).
\end{split}
\end{equation}
Similarly to the two output router, we require that $|J^z| \gg |J^x|$ in order to suppress the state transfer transfer to the wrong output qubits. Depending on the detunings $\Delta_i$, the router can realize different behaviors.

The first setting is realized by tuning $\Delta_1 = \Delta_3 = -J^z$ and $\Delta_2 = 2J^z$. The input state is then sent to qubit 1 if the control state is $\left| 01 \right.\rangle$, to qubit 2 if the control state is $\left| 11 \right.\rangle$, or to qubit 3 if the control state is $\left| 11 \right.\rangle $. For all cases, the transfer time is $T = \pi/( 2 J^x )$. For the $\left| 00 \right.\rangle$ control state both output 1 and 3 are open, and the transfer time is also different at $T' = T/\sqrt{2}$. We assume that suck a configuration does not occur, or is forbidden.

It is also possible to set up the router such that only one output is closed. The input state is then sent to both open output qubits and we thus get an entangled state between the open output qubits. Which qubits become entangled is determined by the control qubits. We can thus produce a quantum controlled entanglement distribution system. For this configuration we require the detunings $\Delta_1 = \Delta_3 = J^z$ and $\Delta_2=0$. When the controls are in state $\left| 11 \right.\rangle$ the input state is transfered to qubit 1 and 3,  for $\left| 10 \right.\rangle$ the state is transfered to qubit 1 and 2, and for $\left| 10 \right.\rangle$ the state is transfered to qubit 2 and 3. The transfer time is $T'=T/\sqrt{2}$ for all cases. In the final control configuration $\left| 00 \right.\rangle$ all of the output qubits are closed and the input state remains in the input qubit.


%